\documentclass[
aps,
prb,
reprint,
showkeys,
twocolumn,
nofootinbib,
]{revtex4-2}

\usepackage{easyReview}
\usepackage{upgreek}
\usepackage{natbib}
\usepackage{amssymb}
\usepackage{color}
\usepackage{amsfonts}
\usepackage{textcomp}
\usepackage{amsmath}
\usepackage[charter]{mathdesign}
\usepackage{tikz}
\newcommand{\ped}[1]{\ensuremath{_{\rm #1}}}
\newcommand{\apex}[1]{\ensuremath{^{\rm #1}}}

\definecolor{link}{RGB}{57,106,177}
\usepackage{float}
\usepackage[caption = false]{subfig}
\usepackage{graphicx}
\usepackage{hyperref}
\hypersetup{
	colorlinks=true,
	linkcolor=link,
	citecolor=link,
	urlcolor=link
}

\begin{document}
	
\title{Anomalous metallic phase in molybdenum disulphide induced\\ via gate-driven organic ion intercalation}
	
\author{Erik Piatti}
\email{erik.piatti@polito.it}
\affiliation{Department of Applied Science and Technology, Politecnico di Torino, I-10129 Torino, Italy}
\author{Jessica Montagna Bozzone}
\affiliation{Department of Applied Science and Technology, Politecnico di Torino, I-10129 Torino, Italy}
\author{Dario Daghero}
\affiliation{Department of Applied Science and Technology, Politecnico di Torino, I-10129 Torino, Italy}

\begin{abstract}
Transition metal dichalcogenides exhibit rich phase diagrams dominated by the interplay of superconductivity and charge density waves, which often result in anomalies in the electric transport properties. Here, we employ the ionic gating technique to realize a tunable, non-volatile organic ion intercalation in bulk single crystals of molybdenum disulphide (MoS\ped{2}). 
We demonstrate that this gate-driven organic ion intercalation induces a strong electron doping in the system without changing the pristine $2H$ crystal symmetry and triggers the emergence of a re-entrant insulator-to-metal transition. We show that the gate-induced metallic state exhibits clear anomalies in the temperature dependence of the resistivity with a natural explanation as signatures of the development of a charge-density wave phase which was previously observed in alkali-intercalated MoS\ped{2}. The relatively large temperature at which the anomalies are observed ($\sim$$150$\,K), combined with the absence of any sign of doping-induced superconductivity down to $\sim$$3$ K, suggests that the two phases might be competing with each other to determine the electronic ground state of electron-doped MoS\ped{2}.

\bigskip
\textbf{Cite this article as:} 

Piatti, E.; Montagna Bozzone, J.; Daghero, D. Anomalous metallic phase in molybdenum disulphide induced via gate-driven organic ion intercalation. \textit{Nanomaterials} \textbf{2022}, \textit{12}, 1842.
%
%

\textbf{DOI:} \href{https://doi.org/10.3390/nano12111842}{10.3390/nano12111842}
\end{abstract}


\maketitle

\section{Introduction}

Transition metal dichalcogenides (TMDs) are layered compounds where the combination of relatively simple crystal structures, complex electronic structures, and electronic correlations give rise to very rich phase diagrams\,\cite{KlemmReview2015}, making them highly promising for fundamental investigations and potential applications alike\,\cite{ManzeliReview2017, ChoiReview2017}. Hexagonal molybdenum disulphide (2$H$-MoS\ped{2}) has been one of the most studied semiconducting TMDs owing to its chemical stability, ease of exfoliation, and indirect-to-direct band gap transition in the monolayer\,\cite{MakPRL2010, SplendianiNL2010, WangNatNano2012}, which make it ideally suited to the development of (opto-)electronic applications\,\cite{SchaibleyReview2016, FerrariNS2015, MakNatPhot2016}. Upon electron doping, 2$H$-MoS\ped{2} displays a complex phase diagram 
as the Fermi level is tuned across its multi-valley electronic band structure\,\cite{GePRB2013, PiattiNL2018, PiattiJPCM2019, SohierPRX2019, GarciaCommPhys2019, RomaninJAP2020, GarciaPRB2020, PimentaAN2022}. This exploration has chiefly been enabled by the development of the ionic gating technique\,\cite{FujimotoReview2013, UenoReview2014, LiuReview2022}
, which exploits the ultrahigh electric fields attainable at a voltage-polarized electrolyte/electrode interface to control the electron and/or hole doping of a material
either via simple electrostatic charge induction\,\cite{YeScience2012, PiattiNE2021, YeNatMater2010, JoNL2015, PiattiPRB2017, XiPRL2016, PiattiLTP2019, LiNature2016, LeiPRL2016, PiattiEPJ2019, WangNatNano2018, PiattiApSuSc2020,  ShiSciRep2015} or more complex electrochemical effects\,\cite{YuNatNano2015, ShiSciRep2015, ShiogaiNatPhys2016, PiattiAPL2017, PiattiApSuSc2018, YingPRL2018, JeongScience2013, LiNL2013, ZhangAN2017, WeiPRB2021, PetachPRB2014, WangNJP2019, ZakhidovAN2020}.

Upon increasing electron doping, 2$H$-MoS\ped{2} undergoes first an insulator-to-metal transition\,\cite{RadisavljevicNatMater2013, WuNatCommun2016, PiattiNatElectron2021} and, at higher doping levels, a metal-to superconductor transition; the superconducting phase, attained both via electrostatic carrier accumulation at the surface\,\cite{YeScience2012, BiscarasNatComm2015, PiattiNL2018, CostanzoNatNano2018} or electrochemical ion intercalation in the bulk\,\cite{ZhangNanoLett2016, PiattiAPL2017, BinSubhan2021}, displays a maximum transition temperature $T\ped{c}\approx{11}$\,K. The highest doping levels also destabilize the 2$H$-MoS\ped{2} crystal structure, promoting the development of charge density wave (CDW) phases\,\cite{EnyashinCTC2012, ZhuangPRB2017, ChenCSB2013, RosnerPRB2014, PiattiApSuSc2018, BinSubhan2021} and/or structural transitions to other polytypes\,\cite{EnyashinCTC2012, ChenCSB2013, GuoNL2015, VoiryCSR2015, LengACSNano2016, XiaNS2017, ZhuangPRB2017}.

Although in most cases the effects of ionic gating vanish when the gate voltage is removed, it has recently been shown that \textit{non-volatile} charge doping of macroscopic bulk specimens of layered crystals is attainable via gate-driven intercalation of either hydrogen\,\cite{BoeriJPCM2021, CuiCPL2019, RafiqueNanoLett2019, MengPRB2022} or organic ions\,\cite{WangCPB2021}. The latter has been proved to be a very effective tool to tune the electronic ground state of layered FeSe\,\,\cite{WangCPB2021} and to boost its superconducting transition temperature\,\cite{ShiNJP2018, XuTCC2022} well beyond that achieved via other means of bulk electron doping. These results make the investigation of the effects of gate-driven organic-ion intercalation in other layered materials highly desirable. 

In this work, we employ the ionic gating technique to intercalate organic ions in bulk $2H$-MoS\ped{2} single crystals at room temperature. We find that the gate-induced intercalation is partially non-volatile and can be tuned by varying the amount of time during which the fully intercalated samples are kept in the ionic liquid after removal of the gate voltage. Our organic ion-intercalated MoS\ped{2} crystals retain the $2H$ crystal symmetry and undergo a doping-induced insulator-to-metal transition, with clear anomalies emerging in the temperature-dependence of the resistivity around $\sim 150$\,K; all samples show no evidence of superconductivity down to $\sim 3$\,K, but an incipient re-entrant transition to an insulating state is observed at the highest doping levels. We discuss how all the observed features can be naturally ascribed to the development of a doping-induced CDW phase assisted by disorder-induced strong localization, which may be in competition with the otherwise expected superconducting phase.

\section{Materials and Methods}

\subsection{Electric-field-driven ion intercalation}

Freshly-cleaved 2$H$-MoS\ped{2} crystals (SPI supplies; typical size 2\,mm$\times$1\,mm$\times$50\,$\upmu$m) were electrically contacted via drop-casted spots of silver paste (RS Components) and immersed in a Duran crucible filled with 1-ethyl-3-methylimidazolium tetrafluoroborate ionic liquid (EMIM-BF\ped{4}, Sigma Aldrich) together with a platinum (Pt) counter electrode. The setup is sketched in Figure~\ref{fig1}a. The gate voltage $V\ped{G}$ between the Pt electrode and the crystal was applied by an Agilent B2961 power source, at room temperature and in ambient atmosphere. The resistivity $\rho$ was monitored in situ by the standard four-wire method. A constant current $I\ped{DS}\approx 1\,\upmu$A, flowing between the drain (D) and source (S) contacts, was supplied by a Keithley 220 current source, and the longitudinal voltage drop $V\ped{xx}$ between the inner voltage contacts was measured by a HP3457 multimeter. The resistivity was then determined as $\rho = V\ped{xx} \,I\ped{DS}^{-1} \,t w\, l^{-1}$, where $t$ and $w$ are the sample thickness and width, and $l$ is the distance between the inner voltage contacts. Common-mode offsets were removed using the current-reversal method. 
Open-circuit 
(OC) conditions were obtained by physically disconnecting the gate electrode from the power source. Ex-situ characterizations were performed after extracting the intercalated MoS\ped{2} samples from the cell and rinsing them thoroughly with acetone and ethanol to remove ionic-liquid residues. The samples were stored in standard desiccators either in argon atmosphere or under low vacuum to avoid moisture contamination.

\subsection{Vibrational spectroscopies}

Both Raman and Fourier-transform infrared (FT-IR) spectra were measured on freshly-cleaved surfaces in ambient conditions. Raman spectra were acquired using a Renishaw InVia H43662 micro-Raman spectrometer. All spectra were acquired using an excitation wavelength of 514\,nm, a laser power $<$\,1\,mW focused through a 100X objective, an exposure time of 20\,s, and 50 accumulations. FT-IR spectra were measured using a Thermoscientific Nicolet 5700 spectrometer in attenuated total reflectance (ATR) mode (Thermoscientific Smartorbit module) in the range between 900 and 3500\,cm\apex{-1}. The spectrum of the EMIM-BF\ped{4} ionic liquid was measured by mixing the liquid with potassium bromide. The determination of the peak areas in the bands centered around $\sim1570$ and $\sim1045$\,cm\apex{-1} in both the liquid and the intercalated MoS\ped{2} samples was performed by fitting the experimental spectra in the two regions to one and four Voigt peak functions respectively, in accordance with the peak assignments.

\subsection{Nano-infrared microscopy}

Scanning probe microscopy images were acquired on freshly-cleaved surfaces in ambient conditions by means of a Bruker Anasys nanoIR3-s atomic force microscope (AFM) combined with a multichip tunable quantum cascade laser (QCL; MIRcat-QT Daylight Solutions) covering the IR spectral ranges 1900--1350\,cm\apex{-1} and 1150--900\,cm\apex{-1}, using commercial gold-coated silicon tips (Bruker PR-EX-TnIR-A). AFM topography maps were acquired in tapping mode, and the IR absorption maps were simultaneously recorded via heterodyne detection as implemented in the ``Tapping AFM-IR" mode of the built-in Analysis Studio software. All maps were acquired with $\sim60$\,\% of the average QCL laser power (0.5\,W), a duty cycle of 15\%, and a pulse rate $\sim300$\,kHz set as to allow for heterodyne detection of the IR signal amplitude. Absorption spectra were collected with a spectral resolution of 2\,cm\apex{-1} and 64 co-averages, and normalised to the QCL emission profile at the same laser power. Gwyddion software\,\cite{Gwyddion} was employed for image analysis.

\begin{figure*}[]
	\includegraphics[width=\textwidth]{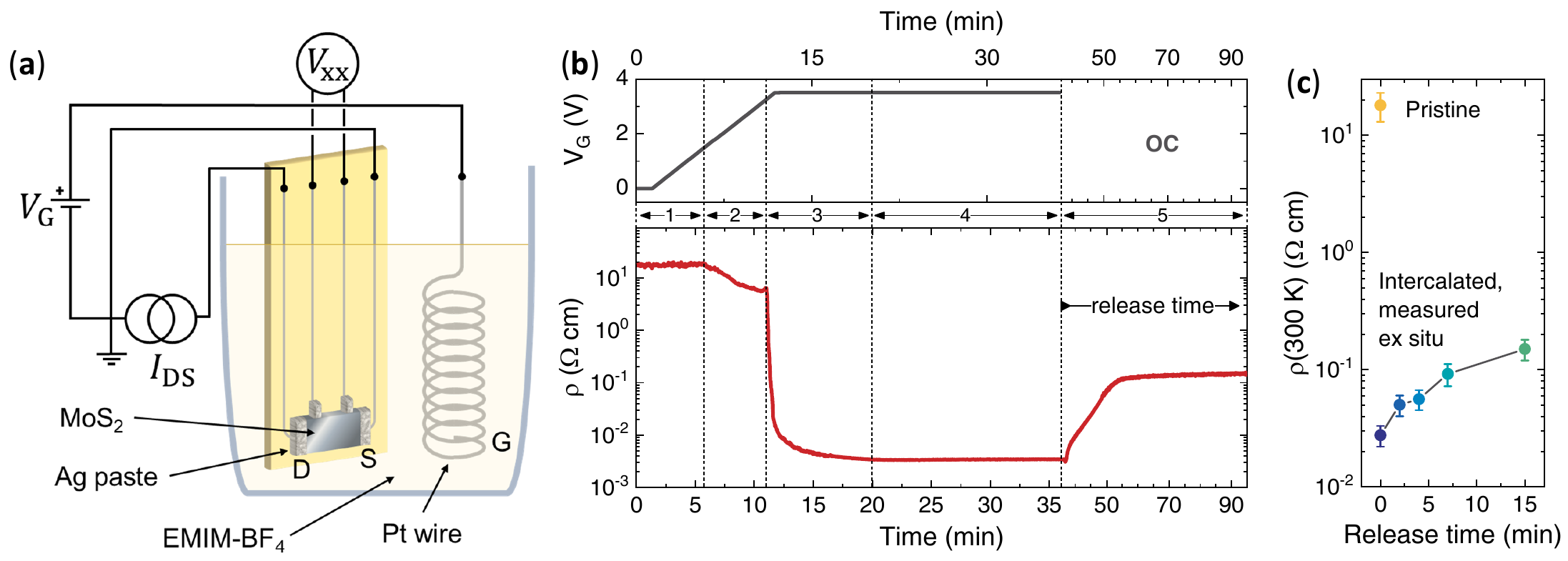}
	\caption{
		Electric-field-driven ion intercalation. (\textbf{a}) Sketch of the ionic gating setup with the electrical connections. Source (S), drain (D) and gate (G) electrodes are explicitly highlighted. (\textbf{b}) Gate voltage (top panel) and resistivity (bottom panel) measured \textit{in situ} as a function of time during a gating process. (\textbf{c}) Resistivity measured \textit{ex situ} as a function of the release time.
		\label{fig1}
	}
\end{figure*}   

\subsection{Kelvin-probe force microscopy}

Maps of the surface potential were acquired on freshly-cleaved surfaces in ambient conditions by means of a Bruker Innova AFM equipped with a surface potential imaging add-on, using standard platinum-iridium-covered silicon tips (Bruker SCM-PIT-V2). Pristine and intercalated samples were mounted simultaneously on the same metallic sample holder and fixed with electrically-conducting adhesive tape. AFM topography maps were acquired in tapping mode during the forward scans, whereas Kelvin-probe force microscopy (KPFM) measurements were performed in lift mode during the backward scan by grounding the sample and biasing the KPFM tip. The $20\times20\,\upmu$m$^2$ maps shown in the following were acquired using a scan rate of 0.3\,Hz, a lift height of 50\,nm, and a tip AC bias voltage of 3\,V. Gwyddion software\,\cite{Gwyddion} was employed for image analysis.

\subsection{Resistivity measurements}

The measurements of the resistivity as a function of temperature were performed in the high-vacuum chamber of a Cryomech pulse-tube cryocooler with a base temperature of $\approx 2.8\,$K. The resistivity was determined with the same method employed in the room-temperature gating runs, except that $I\ped{DS}$ was supplied by an Agilent B2912 source-measure unit, and $V\ped{xx}$ was measured by an Agilent 34420 nanovoltmeter. In each cooling-warming cycle, only the data recorded during the slower, quasi-static warming to room temperature were considered.

\section{Results}

The ionic gating technique is widely employed to control the electronic properties of various classes of materials. The principle of the technique is that the material under study constitutes the active channel of an electrochemical transistor, and is thus separated from a gate counter-electrode by an electrolyte such as an ionic liquid. When polarized by the application of a finite $V\ped{G}$, the electrolyte allows the mobile ions to accumulate at the sample surface and induce an equal and opposite sheet carrier density in the channel to maintain charge neutrality, building up the electric double layer (EDL) which acts as a nanoscale capacitor with ultra-high capacitance\,\cite{FujimotoReview2013, UenoReview2014, LiuReview2022}. The intense electric field in the EDL can then be directly exploited to electrostatically tune this surface charge density as in a conventional field-effect transistor\,\cite{YeScience2012, PiattiNE2021, YeNatMater2010, JoNL2015, PiattiPRB2017, XiPRL2016, PiattiLTP2019, LiNature2016, LeiPRL2016, PiattiEPJ2019, WangNatNano2018, PiattiApSuSc2020,  ShiSciRep2015}, or to activate more complex electrochemical effects including bulk ion intercalation\,\cite{YuNatNano2015, ShiSciRep2015, ShiogaiNatPhys2016, PiattiAPL2017, PiattiApSuSc2018, YingPRL2018, JeongScience2013, LiNL2013, ZhangAN2017, WeiPRB2021, PetachPRB2014, WangNJP2019, ZakhidovAN2020}. Here, we make use of the ionic gating setup sketched in Fig.\,\ref{fig1}a to intercalate positively-charged organic ions in bulk $2H$-MoS\ped{2} crystals and induce a semi-permanent electron doping.

Figure~\ref{fig1}b shows the resistivity $\rho$ of a representative MoS\ped{2} crystal during the gating process at room temperature. At $V\ped{G} = 0$, a finite $\rho\approx 18\,\Omega$\,cm is measured due to the intrinsic $n$-doping arising from the sulphur vacancies ubiquitous in MoS\ped{2} crystals\,\cite{QiuNatCommun2013}. As $V\ped{G}$ is swept from $0$ to positive values, $\rho$ remains mostly unaffected as long as $V\ped{G}\lesssim +1$\,V (region 1 in Figure~\ref{fig1}b) and then starts to decrease, reaching a plateau around $\rho\approx 5.5\,\Omega$\,cm in the $V\ped{G}$ range between $\sim +2.75$ and $\sim+3.25$\,V (region 2 in Figure~\ref{fig1}b). Further increasing $V\ped{G}$ to $+3.5$\,V then triggers a sharp decrease in $\rho$ of about 3 orders of magnitude. If the gate voltage is then kept fixed at $V\ped{G}=+3.5$\,V the resistivity saturates rather quickly ($\lesssim 10$\,min, region 3 in Figure~\ref{fig1}b) to $\rho\approx3$\,m$\Omega$\,cm and shows no further change (region 4 in Figure~\ref{fig1}b). This gate-induced change in $\rho$ is partially volatile: when the potential difference between the Pt counter electrode and the sample is removed by setting the cell in the OC conditions (region 5 in Figure~\ref{fig1}b), $\rho$ rapidly increases, saturating to an intermediate value $\rho\approx150$\,m$\Omega$\,cm after about 15 minutes; however, smaller values of $\rho$ can be stabilized by extracting the MoS\ped{2} crystal from the cell before the full OC saturation has taken place. Figure~\ref{fig1}c shows the $\rho$ of gated MoS\ped{2} crystals measured \textit{ex situ} as a function of the release time, i.e. the time span in which each crystal has been left in the ionic liquid in OC conditions. While the lowest-resistivity state ($\rho\approx3$\,m$\Omega$\,cm) observed in region 4 goes lost as soon as the gate circuit is opened, permanent modifications of the intercalated crystals leading to intermediate $\rho$ values ranging from $\approx 25$\,m$\Omega$\,cm up to $150$\,m$\Omega$\,cm (and monotonically depending on the release time) can thus be obtained. With the exception of the electric transport measurements, all characterizations have been performed on samples at zero (nominal) release time.

\begin{figure*}[]
	\includegraphics[width=\textwidth]{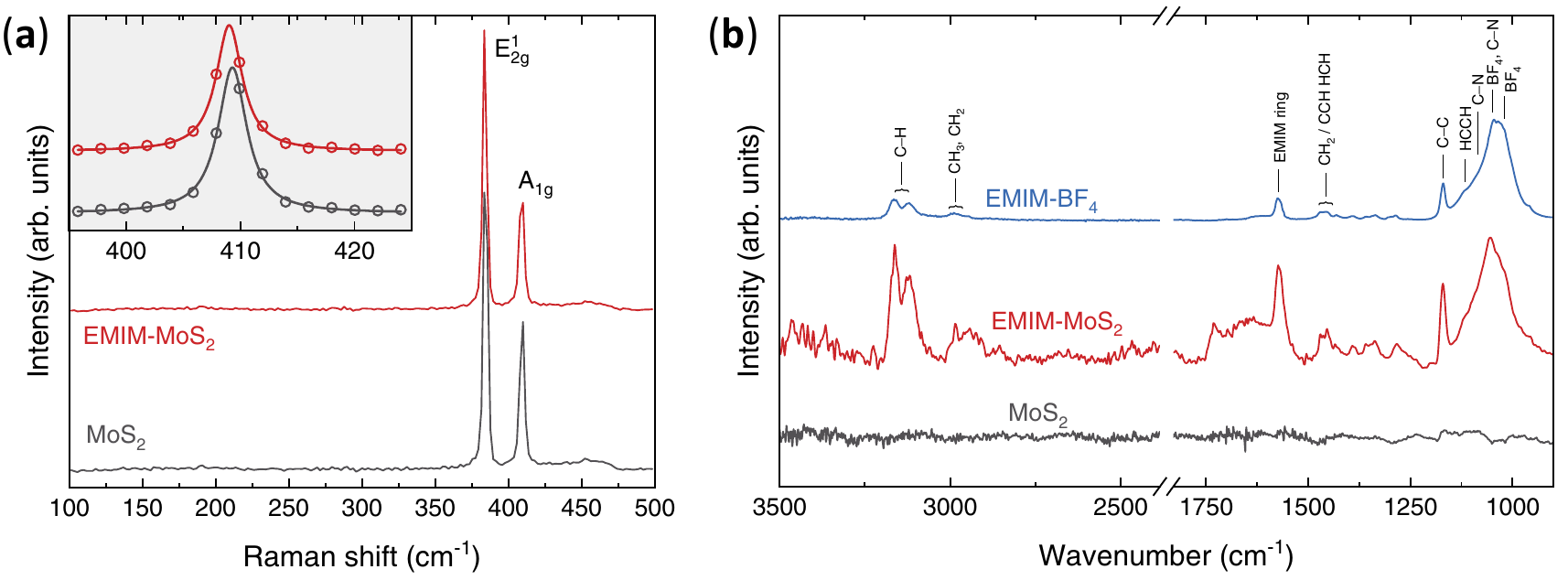}
	\caption{
		Vibrational spectroscopies of pristine and intercalated MoS\ped{2}. (\textbf{a}) Raman spectroscopy at 514\,nm in the spectral range between 100 and 500\,cm\apex{-1}, showing the E\apex{1}\ped{2g} mode at $\sim 384$\,cm\apex{-1} and the A\ped{1g} mode at $\sim 409$\,cm\apex{-1}. Inset shows a magnification of the data in the A\ped{1g} region close to 409\,cm\apex{-1}. Solid lines are Lorentzian fits. (\textbf{b}) FT-IR absorption spectra acquired in ATR mode between 3500 and 900\,cm\apex{-1}. Peak assignments for the spectrum of EMIM-BF\ped{4} are indicated.
		\label{fig2}
	}
\end{figure*}   

Since 2$H$-MoS\ped{2} is prone to undergo structural transitions upon heavy electron doping\,\cite{RosnerPRB2014, ZhuangPRB2017} and intercalation by different ionic species\,\cite{EnyashinCTC2012, ChenCSB2013, GuoNL2015, VoiryCSR2015, LengACSNano2016, XiaNS2017}, the structural phase of both the pristine and the intercalated crystals was assessed via Raman spectroscopy. As shown in Figure~\ref{fig2}a, the Raman spectrum of a pristine crystal (solid black line) exhibits the two typical modes of MoS\ped{2} in the $2H$ crystal structure, the E\apex{1}\ped{2g} mode at $\sim 384$\,cm\apex{-1} and the A\ped{1g} mode at $\sim 409$\,cm\apex{-1}, which correspond to in-plane and out-of-plane vibrations of Mo and S atoms respectively\,\cite{LeeACSN2010, PiattiNL2018}. The same two peaks are also observed in the Raman spectrum of intercalated MoS\ped{2} (solid red line), with no sign of the emergence of the peaks at 158, 218 and 334\,cm\apex{-1} which are associated to the 1$T$/1$T$' MoS\ped{2} polytypes\,\cite{GuoNL2015, LengACSNano2016, VoiryCSR2015, XiaNS2017}. This indicates that our MoS\ped{2} samples retain their $2H$ crystal structure upon EMIM intercalation. Furthermore, Lorentzian fitting to the experimental spectra (inset to Figure~\ref{fig2}a) indicates that the A\ped{1g} mode undergoes a tiny $\approx 0.3$\,cm\apex{-1} redshift in the intercalated sample, whereas the E\apex{1}\ped{2g} mode remains unchanged. Both features are consistent with the behavior of MoS\ped{2} nanolayers upon gate-induced electron doping\,\cite{SohierPRX2019, ChakrabortyPRB2012, LuSmall2017}, even though a quantitative comparison is prevented by the strong dependence of the effect on the sample thickness.

The successful incorporation of EMIM ions in the MoS\ped{2} lattice was confirmed by FT-IR measurements. As shown in Figure~\ref{fig2}b, the infrared spectrum of the pristine MoS\ped{2} crystal (solid black line) is mostly featureless in the investigated frequency ranges, since the sharp infrared peaks typical of the 2$H$ phase emerge at wavenumbers below $\sim500$\,cm\apex{-1}\,\cite{ChenAC2016}. Conversely, the spectrum of the intercalated MoS\ped{2} sample (solid red line) exhibits several bands, that are actually typical of organic compounds. Indeed, they fall in the C--H stretching region (between 3200 and 2800\,cm\apex{-1}), and in the fingerprint region (between 1800 and 900\,cm\apex{-1}). Notably, the same bands are also observed in the spectrum of the EMIM-BF\ped{4} ionic liquid (solid blue line). In accordance with the literature, we assign the structure between 3200 and 3050\,cm\apex{-1} to the C--H stretching modes involving unsaturated C atoms in the EMIM ring
\,\cite{HofftLangmuir2008, BoumedieneJMS2019, HeimerJML2006}; the band between 3000 and 2850\,cm\apex{-1} to the aliphatic C--H modes of the CH\ped{3} group bonded to the EMIM ring, the CH\ped{3} terminal group of the ethyl chain, and the CH\ped{2} group\,\cite{HofftLangmuir2008, BoumedieneJMS2019, HeimerJML2006}; the peak around 1572\,cm\apex{-1} to the EMIM-ring breathing\,\cite{HofftLangmuir2008, BoumedieneJMS2019, HeimerJML2006}; the band around 1450\,cm\apex{-1} to the 
(CH\ped{2})/CCH HCH bending modes\,\cite{HofftLangmuir2008, BoumedieneJMS2019, HeimerJML2006}; the peak around 1170\,cm\apex{-1} to the 
C--C aliphatic stretching modes\,\cite{HofftLangmuir2008, BoumedieneJMS2019, HeimerJML2006}. All these bands are associated purely to the EMIM cation. Conversely, the broad band between 1150 and 900\,cm\apex{-1} comprises at least four peaks arising both from the cation and the BF\ped{4} anion. Specifically, we assign 
the peak around 1116\,cm\apex{-1} to ring HCCH symmetric bending\,\cite{BoumedieneJMS2019, HeimerJML2006};
the peak around 1085\,cm\apex{-1} to C--N stretching\,\cite{BoumedieneJMS2019};
the peak around 1050\,cm\apex{-1} to a combination of BF\ped{4} asymmetric stretching and C--N symmetric stretching\,\cite{HeimerJML2006};
and the peak around 1021\,cm\apex{-1} to BF\ped{4} asymmetric stretching\,\cite{BoumedieneJMS2019, HeimerJML2006}. As such, only this last peak can be uniquely associated to the presence of the BF\ped{4} anion. 

The fact that all the main absorption bands of the EMIM-BF\ped{4} ionic liquid are observed also in the infrared spectrum of the intercalated MoS\ped{2} sample indicates that both the EMIM cation and the BF\ped{4} anion are incorporated in the MoS\ped{2} lattice. However, the relative intensity of the BF\ped{4}-related bands with respect to that of the EMIM-related ones is much lower in the intercalated MoS\ped{2} sample than in the ionic liquid. Specifically, the peak-area ratio between the BF\ped{4}-stretching mode at $\sim 1021$\,cm\apex{-1} and the EMIM-breathing mode at $\sim 1572$\,cm\apex{-1} in the EMIM-BF\ped{4} ionic liquid ($A\ped{BF\ped{4}}/A\ped{EMIM}\approx 12$) is about 10 times larger than that in the intercalated MoS\ped{2} sample ($A\ped{BF\ped{4}}/A\ped{EMIM}\approx 1.2$). Considering that the ionic liquid has a 1:1 EMIM:BF\ped{4} stoichiometry, this allows us to estimate that in the intercalated MoS\ped{2} samples about one BF\ped{4} anion every $\sim10$ EMIM cations is incorporated in the lattice as a contaminant. The fact that the MoS\ped{2} lattice incorporates a much larger amount of cations than of anions confirms a net electron doping in the intercalated samples.

\begin{figure*}[]
\includegraphics[width=0.8\textwidth]{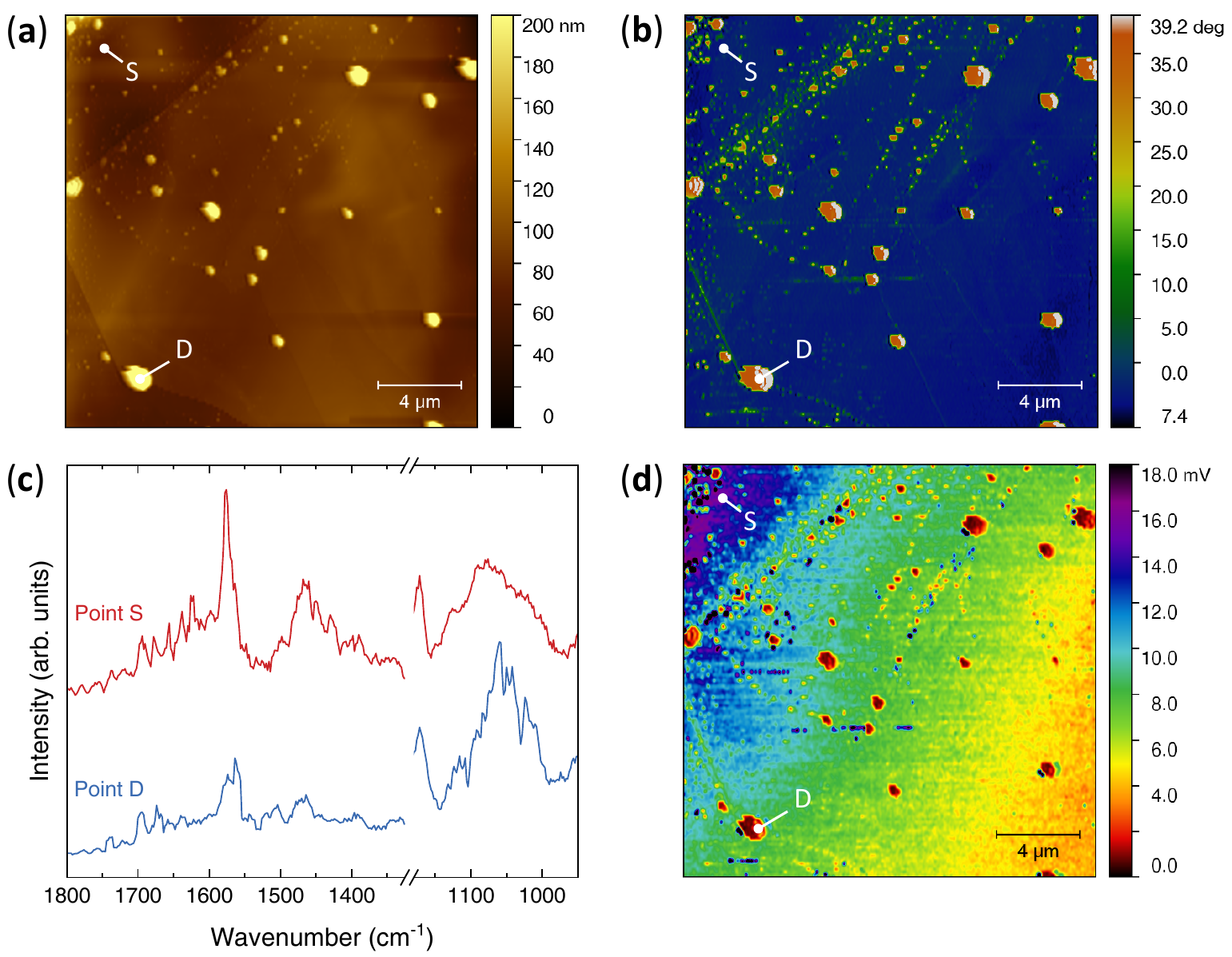}
\caption{
Infrared Nanospectroscopy of intercalated MoS\ped{2}. (\textbf{a}) AFM topography map. (\textbf{b}) Phase image of the same area. (\textbf{c}) nano-IR absorption spectra acquired in correspondence of a droplet (solid blue line, point D) and of the MoS\ped{2} exposed surface (solid red line, point S). (\textbf{d}) IR absorption map of the same area acquired in correspondence of the EMIM-breathing mode at $\sim 1572$\,cm\apex{-1}.
\label{fig3}
}
\end{figure*}

The spatial distribution of the intercalants was investigated by nanoscale infrared spectroscopy\,\cite{PraterMT2010, DazziCR2017}. Figure~\ref{fig3}a shows a $20\times20\,\upmu$m\apex{2} AFM topography map of a freshly-cleaved MoS\ped{2} crystal. Unlike the nearly atomically-flat surface of pristine crystals (discussed later), intercalated MoS\ped{2} exhibits a somewhat corrugate surface (RMS roughness $S\ped{q}\approx15$\,nm) although atomically-flat terraces can still be observed. Furthermore, much taller ($\approx 200$\,nm) ellipsoidal features  are randomly scattered on top of the surface and exhibit a very strong contrast in the tapping phase image (Figure~\ref{fig3}b), which indicates that their mechanical properties differ significantly from those of the MoS\ped{2} background. IR absorption spectra were then collected both on one of these "droplets" (point D) and on the exposed surface (point S). As shown in Figure~\ref{fig3}b, both spectra display the peaks, specific of the EMIM cation, at about $1572$\,cm\apex{-1}, $1450$\,cm\apex{-1}, and $1170$\,cm\apex{-1}, as well as the broad band between 1150 and 900\,cm\apex{-1}, that also includes the BF$_4$-related peak at $\sim 1021\,\mathrm{cm}^{-1}$. However, in the spectrum acquired on the droplet (solid blue line) this latter band is much more pronounced than the EMIM-related peaks, as it happens in the FT-IR spectrum of the pure ionic liquid shown in Figure~\ref{fig2}b. Conversely, in the spectrum acquired on the exposed MoS\ped{2} surface, the BF\ped{4}-related band is significantly less intense with respect to the EMIM-related peaks: in particular, the spectral weight of the BF\ped{4}-stretching peak at $\sim1021$\,cm\apex{-1} is strongly suppressed. Combined with the phase contrast, these results strongly suggest that the droplets are due to ionic-liquid residues embedded within the MoS\ped{2} planes (and exposed to the surface after cleaving), and their presence is the main source of BF\ped{4} anions detected by FT-IR. Finally, the spatial distribution of the intercalated EMIM cations was studied by acquiring an IR absorption map in correspondence of the EMIM-breathing mode at $\sim 1572$\,cm\apex{-1}: as shown in Figure~\ref{fig3}d, a finite IR absorption at this wavenumber can be observed throughout the entire surface, and it is notably minimum in the areas covered by the ionic-liquid residues; however, a long-range modulation in the IR absorption over a length scale $\sim10\,\upmu$m can be distinguished, which indicates that EMIM incorporation in the MoS\ped{2} lattice is inhomogeneous and develops local minima and maxima over a comparable length scale.

\begin{figure*}[]
\includegraphics[width=\textwidth]{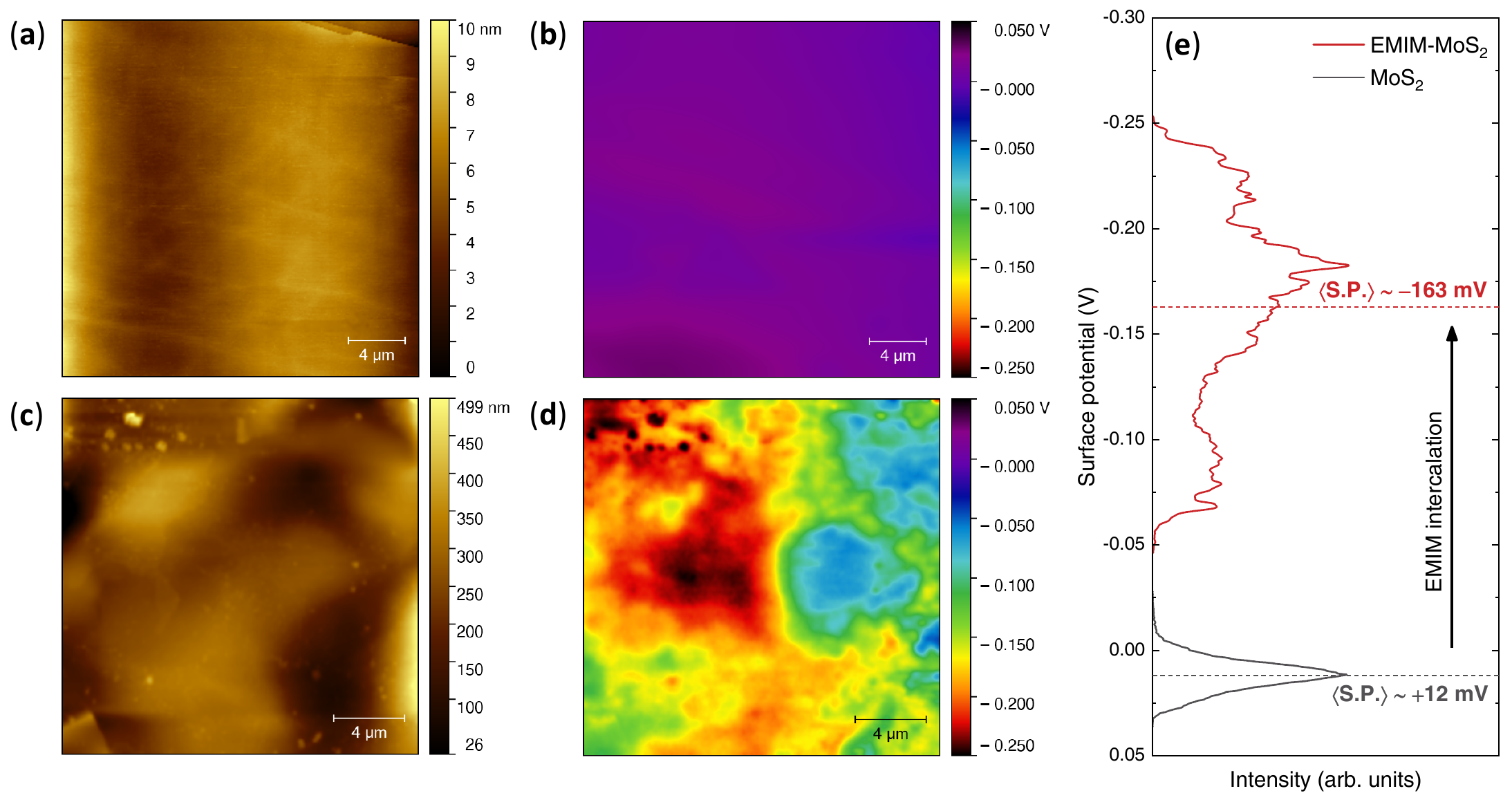}
\caption{
Kelvin-probe force microscopy of pristine and intercalated MoS\ped{2}. (\textbf{a}) AFM topography map of pristine MoS\ped{2}. (\textbf{b}) Surface potential map of pristine MoS\ped{2}. (\textbf{c}) AFM topography map of intercalated MoS\ped{2}. (\textbf{d}) Surface potential map of intercalated MoS\ped{2}. (\textbf{e}) Surface potential histograms highlighting the shift in the Fermi energy induced by the intercalation of EMIM ions.
\label{fig4}
}
\end{figure*}

The impact of the EMIM intercalation on the electronic structure of MoS\ped{2} was assessed by means of Kelvin-probe force microscopy\,\cite{KPFMReview, AlmadoriAAMI2018, DingAPA2019} on a pristine MoS\ped{2} crystal and on the same intercalated crystal characterized via nano-IR. Figure~\ref{fig4}a and b show the AFM topography map and surface potential distribution of the pristine MoS\ped{2} respectively. As expected for the cleaved surface of a layered material, the topography is very flat ($S\ped{q}\approx1.5$\,nm), atomic steps and terraces can be distinguished, and the surface potential is featureless and nearly homogeneous across the entire area. The topography map of intercalated MoS\ped{2} (Figure~\ref{fig4}c) again exhibits a much larger background corrugation ($S\ped{q}\approx15$\,nm) and the localized features associated to ionic-liquid contaminants. More importantly, the surface potential of intercalated MoS\ped{2} (Figure~\ref{fig4}d) is inhomogeneous and exhibits fluctuations over the same length scale ($\sim 10\,\upmu$m) that characterizes the fluctuations in the EMIM content detected via nanoIR. As clearly shown by the direct comparison of the histograms of the two images reported in Figure~\ref{fig4}e, the surface potential of intercalated MoS\ped{2} is also shifted to lower values with respect to that of pristine MoS\ped{2} across the entire area. This reduction in surface potential is a direct proof of electron doping due to the insertion of the EMIM cations in the MoS\ped{2} lattice, since it signifies that in the intercalated MoS\ped{2} the Fermi level has been shifted upwards in energy and closer to the vacuum level. 

The average shift in Fermi energy $\Delta E\ped{F}$ between the two systems can then be estimated simply as:
\begin{equation}
\Delta E\ped{F} = - e\left(\langle \mathrm{S.P.} \rangle\ped{intercalated} - \langle \mathrm{S.P.} \rangle\ped{pristine}\right) = 175\,\mathrm{meV}
\end{equation}
where $e$ is the elementary charge, $\langle \mathrm{S.P.} \rangle\ped{pristine}=+12$\,mV and $\langle \mathrm{S.P.} \rangle\ped{intercalated}=-163$\,mV are the average values of the surface potential in pristine and intercalated MoS\ped{2} respectively. From this shift in the Fermi energy the average density of electrons $n_e$ doped into the conduction band of MoS\ped{2} can also be determined, since the Fermi level of pristine MoS\ped{2} is known to be pinned to the bottom of the conduction band by the presence of sulphur vacancies in the crystal\,\cite{QiuNatCommun2013}. In particular, assuming a parabolic three-dimensional dispersion of the density of states of bulk MoS\ped{2}: 
\begin{equation}
n_e = \frac{\nu_s\nu_v}{6\pi^2} \left(\frac{2m^*}{\hbar^2} \Delta E\ped{F}\right)^\frac{3}{2}
\end{equation}
with a spin degeneracy $\nu_s=2$, a valley degeneracy $\nu_v=6$\,\cite{WuNatCommun2016, ChenPRL2017}, and an effective mass $m^*=0.6 m_e$\,\cite{ChenPRL2017}, one obtains $n_e\sim9\times10^{20}$\,cm\apex{-3}. If every intercalated EMIM cation were to donate one electron to the MoS\ped{2} conduction band, this would correspond to about one EMIM cation every $\sim20$ MoS\ped{2} formula units, or a nominal stoichiometry of EMIM\ped{0.05}MoS\ped{2}.

\begin{figure*}[]
\includegraphics[width=0.8\textwidth]{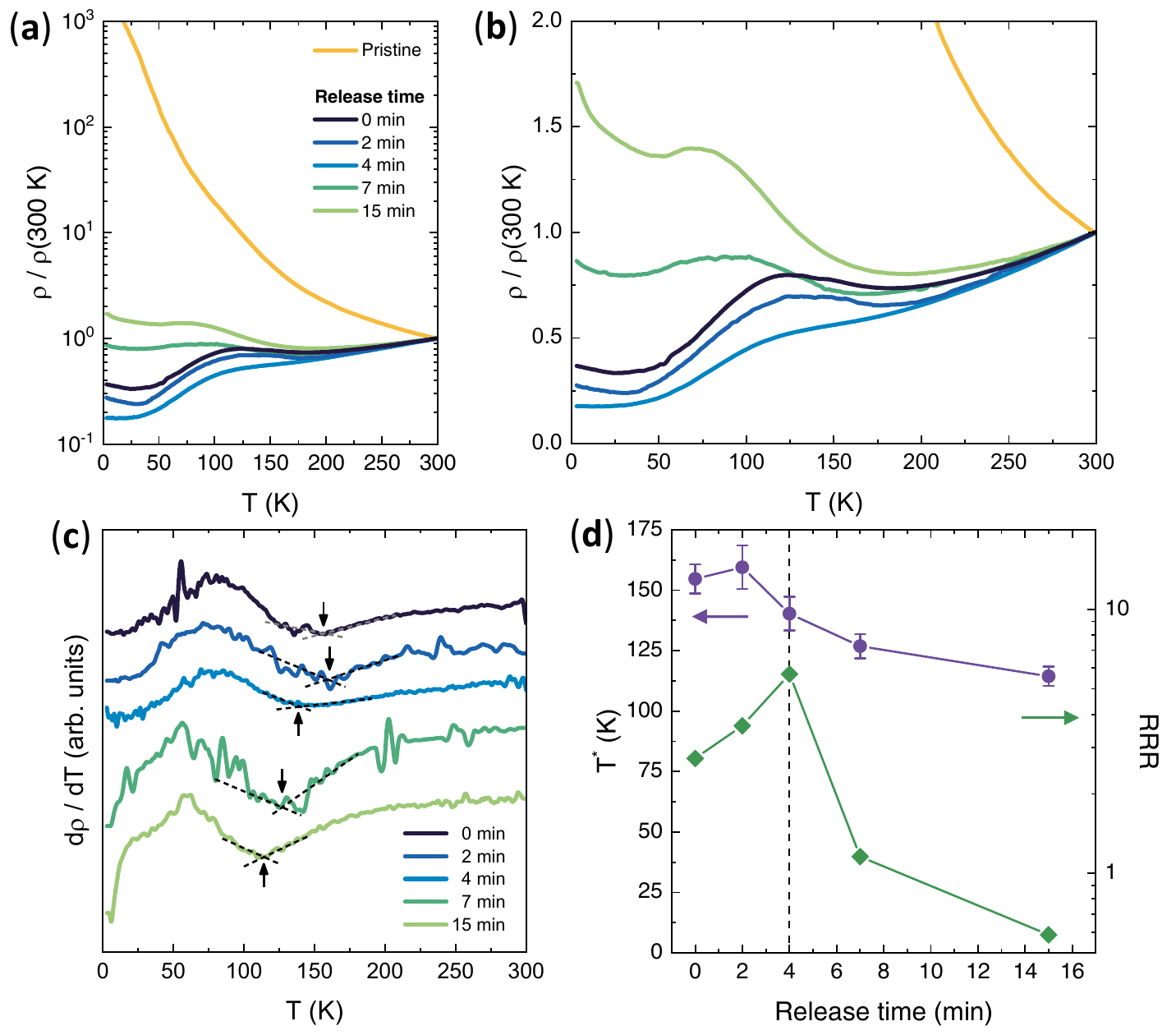}
\caption{
Temperature dependence of the electric transport in pristine and intercalated MoS\ped{2}. (\textbf{a}) Resistivity $\rho$ divided by its value at 300\,K as a function of temperature $T$, before and after EMIM intercalation, in semi-logarithmic scale. (\textbf{b}) same as in (\textbf{a}), but on a linear scale. (\textbf{c}) $T$ dependence of $d\rho/dT$, before and after EMIM intercalation, obtained by numerical derivation of the curves in (\textbf{b}). Arrows highlight the values of $T$ where $d\rho/dT$ shows a dip ($T\apex{*}$). The curves are vertically offset for clarity. (\textbf{d}) $T\apex{*}$ (violet circles, left scale) and residual resistivity ratio $RRR = \rho(300\,\mathrm{K})/\rho(3\,\mathrm{K})$ (green diamonds, right scale) as a function of the release time.
\label{fig5}
}
\end{figure*}   

Having established the mechanism and amount of electron doping provided by the EMIM intercalation in the MoS\ped{2} crystals, we now turn to consider how this affects their electric transport properties as a function of temperature $T$. Figure~\ref{fig5}a shows the $T$-dependence of $\rho$, divided by its value at $T=300$\,K (shown in Figure~\ref{fig1}c), down to $T\sim3$\,K and for a series of MoS\ped{2} crystals with increasing release times. The pristine MoS\ped{2} crystal shows the exponentially-increasing $\rho$ with decreasing $T$ typical of insulators where conduction occurs via hopping processes in the localized states in the band tails, as expected\,\cite{QiuNatCommun2013, PiattiNatElectron2021}.  The $T$-dependence of $\rho$ in the intercalated samples is much less steep, and its values at low $T$ are several orders of magnitude lower than that of pristine MoS\ped{2}, consistent with a metallization induced by the electron doping. A closer inspection of the curves for the intercalated samples (Figure~\ref{fig5}b) reveals a complex dependency of $\rho$ on both $T$ and release time. 
The sample at zero release time  exhibits an overall metallic behavior, in the sense that its $\rho$ decreases on going from 300\,K to 3\,K; however, this $T$-dependence is not monotonic, and a clear anomaly can be observed in the form of a broad hump appearing between 50 and 150\,K. This behavior is best visualized by plotting the first derivative of $\rho$ with respect to $T$, $d\rho/dT$. As shown in Figure~\ref{fig5}c, this anomaly gives rise to a hump-dip structure in the $T$-dependence of $d\rho/dT$, with a broad hump at lower $T$ and a sharp dip at higher $T$, the latter being highlighted by the dashed lines. Additionally, a small resistance upturn is observed for $T\lesssim 20$\,K, indicating that at low $T$ some degree of carrier localization is present in the sample. 

We track the evolution of these two main features -- degree of metallicity and resistivity anomaly -- on increasing the release time, by defining two distinct figures of merit: for the metallicity, the residual resistivity ratio $RRR$, defined as usual as $\rho(300\,\mathrm{K})/\rho(3\,\mathrm{K})$; for the resistivity anomaly, the temperature $T^*$ where the dip in $d\rho/dT$ is located. Figure~\ref{fig5}d shows that, upon increasing the release time (i.e., decreasing the electron doping), $T^*$ (violet circles, left scale) decreases in a nearly monotonic fashion from $\sim 155$\,K at release times below $\sim 2$\,min, to $\sim 115$\,K at release times $\gtrsim 15$\,min. At the same time, however, $RRR$ (green diamonds, right scale) exhibits a clearly non-monotonic trend: after an initial increase from $\sim2.7$ to $\sim5.7$ for release times $\lesssim 4$\,min, it strongly decreases down to $\sim0.58$ at release times $\gtrsim 15$\,min, i.e. when spontaneous deintercalation of the EMIM ions stops. This indicates that the intercalation stage obtained with a release time of 4\,min achieves a maximum degree of metallicity, and doping the sample away from it shifts the system towards an insulating behavior.

\section{Discussion}

In our EMIM-intercalated MoS\ped{2} crystals, increasing the release time reduces the doping and therefore brings the samples back toward the MoS\ped{2} pristine state of a band insulator. On the other side, reducing the release time increases the doping level and brings the samples closer to an insulating state that, however, must have a different origin, since increasing the concentration of EMIM dopants should in principle provide a larger density of free electrons to the conduction band. Such insulating state must therefore be due to some kind of localization. 

A first source of localization could be the randomness caused by intercalation and progressive introduction of disorder in the form of extra scattering centers due to the presence of the ions themselves\,\cite{GallagherNatCommun2015, SaitoACSNano2015, PiattiApSuSc2017, Gonnelli2DMater2017, PiattiAPL2017, PiattiPRM2019}: this type of gate-driven re-entrant transition (from a band insulator, to a metal, and finally to an Anderson insulator) has already been reported in other ion-gated TMDs\,\cite{OvchinnikovNatCommun2016, LuPNAS2017, QinRS2020} and oxides\,\cite{WeiPRB2021}. 
A second source of localization could instead be associated to the same phenomenon giving rise to the resistivity anomaly, which is often found among TMDs and represents one of the typical signatures of the onset of CDW phases\,\cite{KlemmReview2015}. These consist in a periodic modulation of the charge carrier density accompanied by a distortion of the underlying crystal lattice\,\cite{GrunerBook}, usually resulting from strong electron-phonon coupling\,\cite{RossnagelJPCM2011} and/or Fermi-surface nesting\,\cite{ZhuPNAS2015}. CDW phases are very susceptible to charge carrier doping: in most TMD compounds, CDWs are suppressed by increasing doping in favor of the development of superconducting phases\,\cite{LiNature2016, MorosanNatPhys2006, YuNatNano2015, BhoiSciRep2016, FangPRB2005}, but in some specific compounds the CDW order is instead promoted by doping\,\cite{XiPRL2016, PiattiApSuSc2018, BinSubhan2021}. 
Density functional theory calculations have predicted $2H$-MoS\ped{2} to fall into the latter category, both in the bulk\,\cite{EnyashinCTC2012, ChenCSB2013} and in the monolayer\,\cite{RosnerPRB2014, ZhuangPRB2017} form. Recent experimental evidences have confirmed that, indeed, the onset of a CDW order in MoS\ped{2} bulk crystals and bulk-like flakes is promoted by Li-doping\,\cite{PiattiApSuSc2018} and K-doping\,\cite{BinSubhan2021}, as well as by pressure\,\cite{CaoPRB2018}. In this context, $T\apex{*}$ would mark the critical temperature for the CDW phase transition in EMIM-intercalated MoS\ped{2}, placing it in an intermediate range between those of Li\ped{x}MoS\ped{2} ($\approx 230$\,K) and K\ped{x}MoS\ped{2} ($\approx 85$\,K). 

The onset of CDW order can naturally explain why resistivity anomalies emerge in our EMIM-intercalated MoS\ped{2} crystals, and can also provide an additional possible mechanism for the observed incipient localization at high doping, i.e. the opening of a (possibily partial) gap on the Fermi surface associated to the CDW\,\cite{ChenCSB2013, BinSubhan2021}.
It is interesting, at this stage, to understand the reasons why superconductivity is not observed in MoS\ped{2}, at least down to 2.8 K, upon EMIM intercalation. The fact is that both Li-doping and K-doping instead do lead to the appearance of a superconducting phase, although with some differences. 
In K-intercalated MoS\ped{2}, where the CDW order sets in at a low $T^*$, bulk superconductivity occurs with a critical temperature $T\ped{c}\approx 7$\,K\,\cite{ZhangNanoLett2016, BinSubhan2021} throughout most of its phase diagram with little change in $T\ped{c}$ as a function of the K content (between K\ped{0.13}MoS\ped{2} and K\ped{1.5}MoS\ped{2})\,\cite{ZhangNanoLett2016}.
In Li-intercalated MoS\ped{2}, where $T^*$ is high, superconductivity (with $T\ped{c}\approx3.7$\,K) is observed only in samples showing no sign of CDW-related resistivity anomalies\,\cite{PiattiAPL2017}.
In our EMIM-intercalated crystals, in which the CDW order is ubiquitous, but sets in at intermediate temperatures $T^*$, no sign of superconductivity has been detected. One could object that, maybe, the range of doping contents of our samples is not the right one to observe superconductivity. This is certainly possible, but since the EMIM doping distribution is largely inhomogeneous, the system actually experiences a wide range of local electron doping levels, from very small ones to others, much larger than the highest average doping content calculated above. This means that, if there were a sufficiently wide range of doping for superconductivity to develop, at least a fraction of the volume should fall in that range. Even if this fraction were not sufficient to ensure superconducting percolation, a reduction in the resistivity should be observed. The fact that this does not happen suggests that the superconducting phase either does not develop at all or, if it does, has a very low $T_c$ (smaller than 2.8 K). 

This whole picture seems to indicate that the CDW order is competing with superconductivity in MoS\ped{2}, so that the superconducting phase appears with higher $T_c$ only when the CDW is suppressed or at least $T^*$ is lower. 
Alternatively, the lack of superconductivity in EMIM-intercalated MoS\ped{2} could be due to the large size of the organic ions, which strongly increases the interlayer spacing: this would make the intercalated system akin to a stack of electronically-decoupled MoS\ped{2} monolayers, where superconductivity occurs with a much lower $T\ped{c}\lesssim 2$\,K and is much more easily suppressed by disorder\,\cite{CostanzoNatNano2016, FuQM2017}. Such decoupling was reported in the case of another TMD compound when intercalated via organic molecules, SnSe\ped{2}, where however the electronic decoupling was shown to increase the superconducting $T\ped{c}$\,\cite{Wu2DM2019, MaPRM2020}.
Overall, which of the depicted scenarios -- or combination thereof -- is at the origin of the anomalous metallic transport exhibited by our EMIM-intercalated MoS\ped{2} crystals is left as an open question, making future investigations highly desirable. Electric transport measurements below $\sim3$\,K would of course be needed to ascertain whether superconductivity develops at lower temperatures than those achievable in the setup used in the current study. Moreover, low-temperature structure-sensitive techniques such as Raman spectroscopy, X-ray diffraction, and scanning tunnelling microscopy would be highly beneficial to provide direct evidence for CDW ordering in the system, as well as to track its evolution as a function of temperature and doping.

\section{Conclusions}


In summary, we synthesized organic ion-intercalated molybdenum disulphide crystals via the ionic gating method with EMIM-BF\ped{4} ionic liquid. We showed that the intercalation is partially non-volatile and can be tuned by allowing spontaneous de-intercalation to occur when the crystals are left in the electrochemical cell in open-circuit conditions. By combining vibrational spectroscopies, Kelvin-probe force microscopy, and nano-infrared microscopy, we demonstrated that the EMIM ions act as electron donors and are intercalated in the entire sample volume, with their density exhibiting fluctuations over a length scale of the order of $\sim 10\,\upmu$m. We characterized the temperature dependence of the resistivity in the intercalated crystals, showing how increasing EMIM intercalation drives the system across a re-entrant insulator-to-metal transition, i.e. from the pristine band insulator into an anomalous metallic state, and eventually towards an incipient second insulating state. We further revealed the appearance of a doping-induced hump in the resistivity around $\sim 150$\,K, which we attribute to the development of a charge-density wave phase on the basis of the established literature on Li-doped and K-doped MoS\ped{2}. A qualitative comparison between these three systems, in particular focused on the relation between the appearance of superconductivity and the onset of the charge-density-wave phase, is suggestive of a competition between these two quantum phases in electron-doped MoS\ped{2}.

\acknowledgments{
This research was funded by the Italian Ministry of Education, University and Research (Project PRIN Quantum2D, Grant No. 2017Z8TS5B).
We are grateful to M. Bartoli for assistance in the FT-IR interpretation and analysis, and to R. S. Gonnelli for fruitful scientific discussions. The nano-infrared microscopy characterization was conducted at the facilities supported by the project ``Dipartimento d'Eccellenza'' of the Department of Applied Science and Technology at Politecnico di Torino.
}

\bibliographystyle{naturemag}
\bibliography{bibliography}

\end{document}